\documentclass[twocolumn,showpacs,preprintnumbers,prb]{revtex4}
\usepackage{graphicx}
\usepackage{epsfig}
\usepackage{rotating}
\usepackage{dcolumn}
\usepackage{bm}
\usepackage{amssymb,amsmath}
\begin{document}

\title{Kinetic Monte Carlo Simulations of a Model for 
Heat-assisted Magnetization Reversal in Ultrathin Films}


\author{
W.~R.\ Deskins,$^{1,2}$
G.~Brown,$^{1,3}$
S.~H.\ Thompson,$^{1,2}$ 
P.~A.\ Rikvold$^{1,4}$
}\email{prikvold@fsu.edu}

\affiliation{
$^1$Department of Physics, Florida State University,\\ 
Tallahassee, Florida 32306-4350, USA\\
$^2$Department of Scientific Computing, Florida State University,\\ 
Tallahassee, Florida 32306-4120, USA\\
$^3$Computational Science and Mathematics Division, 
Oak Ridge National Laboratory, Oak Ridge, Tennessee 37831-6164, USA\\
$^4$National High Magnetic Field Laboratory, Tallahassee, Florida 32310-3706, USA\\
}
\date{\today}

\begin{abstract}
To develop practically useful systems for ultra-high-density information 
recording 
with densities above terabits/cm$^2$, it is necessary to simultaneously 
achieve  
high thermal stability at room temperature and high recording rates. One method 
that has been proposed to reach this goal is 
{\it heat-assisted magnetization reversal\/} (HAMR). In this method, the 
magnetic orientation is assigned to a high-coercivity material by temporarily 
reducing the coercivity during the writing process through localized heating.
Here we present kinetic Monte Carlo 
simulations of a model of HAMR for ultrathin films, in which the temperature 
in the central part of the film is momentarily increased above the critical 
temperature, for 
example by a laser pulse. We observe that the speed-up achieved by 
this method, relative to the 
switching time at a constant, subcritical temperature, is optimal for an 
intermediate 
strength of the writing field. This effect is explained using the theory of 
nucleation-induced magnetization switching in finite systems. 
Our results should be particularly relevant to 
recording media with strong perpendicular anisotropy, such as 
ultrathin Co/Pt or Co/Pd multilayers. 
\end{abstract}


%
%
%

\maketitle



\section{Introduction}
\label{sec-int}

One of the most important factors supporting progress in the miniaturization of 
computers and other electronic devices is the continued 
exponential increase in the density of data storage.\cite{MCDA05} 
Currently, designs are being considered for magnetic recording devices that 
have areal data densities of the order of terabits/cm$^2$ -- several orders 
of magnitude more than only a decade ago. At
such densities, the size of the recording bit approaches the 
superparamagnetic limit, where thermal fluctuations seriously degrade the 
stability of the magnetization.\cite{BEAN59,RICH94} However, 
current industry standards demand that bits should retain 95\% of their 
magnetization over a period of ten years.\cite{MCDA05} Furthermore, 
subnanosecond magnetization-switching times are required to achieve acceptable
read/write rates. 

One suggested method to fulfill these requirements is to use ultrathin, 
perpendicularly magnetized films of very high-coercivity materials, 
such as FePt (coercive field about 50~kOe), or single-particle bits that are 
expected to have even higher coercivities.\cite{MCDA05}  
However, such high coercive fields at room temperature 
are beyond what is achievable by modern write heads, 
which are limited to about 17~kOe.\cite{MATS06}
A method suggested to overcome this problem is to exploit the temperature 
dependence of the coercivity
through heat-assisted magnetization reversal or HAMR (aka.\ thermally assisted 
magnetization reversal or 
TAMR).\cite{MCDA05,MATS06,SAGA99,KATA00,PURN07,WASE08,PURN09,CHAL09,STIP10,OCON10} 
This is accomplished by increasing the temperature of the recording area
to a value close to, or above, the Curie temperature of the medium via a
localized heat source, such as a 
laser.\cite{MATS06,KATA00,WASE08,CHAL09,STIP10,OCON10} Due to
the temperature dependence of the coercivity, the magnitude of the required 
switching field is lowered at the elevated temperature, relaxing the 
requirements for the write head.
An important consideration for the implementation of the HAMR technique is to 
keep the heat input as low and as tightly focused as possible, 
limiting energy transfer to neighboring recording bits. In order to reach the 
desired high data densities, the laser spot must have a diameter less than 
50~nm, much smaller than the wavelength. 
This can be achieved using near-field optics, a technology which 
currently is the objective of vigorous 
research and development.\cite{MATS06,CHAL09,STIP10,OCON10} 

Despite their simplicity, two-dimensional kinetic Ising models have been 
shown to be useful for studying magnetization switching 
in ultrathin films with strong anisotropy.\cite{RICH94} 
Theoretical \cite{BAND88} and experimental \cite{BACK95}
work has shown that the equilibrium phase transition in such films 
belongs to the universality class of the two-dimensional Ising model. 
The dynamics of magnetization switching in ultrathin, perpendicularly 
magnetized films has been studied using magneto-optical microscopies in 
combination with Monte Carlo simulations of Ising-like models by, 
among others, Kirilyuk et al. \cite{KIRI97} and Robb et al.\cite{ROBB08} 
Systems that have been found to have strong Ising character include 
Fe sesquilayers\cite{BACK95} and ultrathin films of Co,\cite{KIRI97} 
Co/Pd,\cite{CARC85,PURN09} and Co/Pt.\cite{ROBB08,FERR02} 
The strong anisotropy in 
such systems limits the effects of transverse spin dynamics and ensures 
that local spin reversals are thermally activated. The extreme thinness of the 
films strongly reduce the demagnetization effects to which films with 
out-of plane magnetization are otherwise 
subject.\cite{BAND88,BACK95,ROBB08}  
For detailed reviews of experimental and simulational studies of magnetization 
switching in ultrathin films with perpendicular magnetization, see Refs.\ 
\onlinecite{FERR02,LYBE00}. 

In the present paper we use a two-dimensional Ising ferromagnet to model 
the HAMR process by kinetic Monte Carlo (MC) simulation, 
demonstrating enhanced nucleation of the switched magnetization state in the 
heated area. 
For simplicity and computational economy, we envisage an experimental setup 
slightly different from others previously reported in the 
literature.\cite{SAGA99,MATS06,WASE08,PURN09} 
It most closely resembles the optical-dominant setup shown in Fig.~1(b) of 
Ref.~\onlinecite{MATS06}. 
The recording medium is placed in a constant 
write field that is too weak to cause significant switching on an acceptable 
time scale, and it is heated at its center by a transient heat pulse. 
At a fixed superheating temperature we show that the 
relative speed-up of the magnetization switching, compared to the 
constant-temperature case, depends 
nonmonotonically on the magnitude of the applied field. This relative 
speed-up shows a pronounced 
maximum at an intermediate value of the applied field. 
We give a physical explanation for this 
effect, based on the nucleation theory of magnetization switching 
in finite-sized systems.\cite{RICH94,RICH95C,RIKV94A} 
As magnetization switching is a special case of the decay of a metastable 
phase (i.e., the medium in its state of magnetization opposite to the 
applied field),\cite{RIKV94A,RIKV94} 
this analysis is of general physical interest 
beyond the specific technological application discussed here. 

The rest of this paper is organized as follows. 
Our model and methods are described in 
Sec.~\ref{sec-mod}, the numerical results are described 
and explained in Sec.~\ref{sec-res},
and our conclusions are stated in Sec.~\ref{sec-conc}. 
      
\section{Model and Methods}
\label{sec-mod}

We use a square-lattice, nearest-neighbor 
Ising ferromagnet with energy given by the Hamiltonian, 
\begin{equation}
\mathcal{H} = 
-J \sum_{\langle i,j \rangle} s_{i}s_{j} - H \sum_{i} s_{i} \;.
\label{eq:ham}
\end{equation}
Here, $s_i = \pm1$, $J>0$ 
is the strength of the spin interactions, and the first
sum runs over all nearest-neighbor pairs.  For convenience we hereafter set 
$J=1$. In the second term, which represents the Zeeman energy, $H$ is 
proportional to a uniform external magnetic field, and the sum runs over all 
lattice sites. We use a lattice of size $L^{2}=128\times128$,  with periodic 
boundary conditions. 
The length unit used in this study is the computational lattice constant, 
which should correspond to a few nanometers. 

For simplicity, our model does not 
include any explicit randomness, such as impurities or random interaction 
strengths. As a result, pinning of 
interfaces for very weak applied field,\cite{KIRI97,ROBB08} 
as well as heterogeneous nucleation of spin reversal\cite{KIRI97} 
are neglected. We further exclude 
demagnetizing effects, which are very weak for 
ultrathin films\cite{BAND88,BACK95,ROBB08} 
and thus cause no qualitative 
changes in Monte Carlo simulations of the switching process.\cite{RICH95C} 

The stochastic spin dynamic is given by the single-spin flip 
Metropolis algorithm with transition probability\cite{METR53} 
\begin{equation}
P(s_i \rightarrow -s_i) = \min[1, \exp(- \Delta E / T)] \;,
\label{eq:prob}
\end{equation} 
where $\Delta E$ is the energy change that would result from 
acceptance of the proposed 
spin flip. The temperature, $T$, is given in energy units (i.e., Boltzmann's 
constant is taken as unity). Updates are attempted for randomly chosen spins, 
and $L^2$ attempts constitute one MC step per spin (MCSS), which is the time 
unit used in this work. (We note that the Metropolis algorithm is not the 
only Monte Carlo dynamics that could be used here. We have chosen it 
because of its simplicity and ubiquity in the literature since we do not 
expect that the inclusion of complications such 
intrinsic barriers to single-spin flips would have significant effects 
at this high temperature beyond a renormalization of the overall timescale.)  

\begin{figure}[ht]
\centering
\epsfig{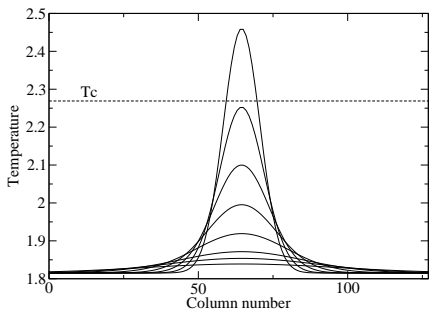} 
\caption{The time dependent Gaussian temperature profile used to simulate 
the decay of a laser heat pulse applied at the center line of the Ising 
lattice.  The times plotted are $t = 1$, 10, 25, 50, 100, 200, 300, and 
500~MCSS.  The tallest Gaussian corresponds to $t=1$~MCSS.}
\label{fig:pulse}
\end{figure}

Following this algorithm and starting from $s_i=-1$ for all $i$,    
we equilibrate the system over $4\times10^{4}$~MCSS at $H=0$ 
and temperature $T_0 = 0.8T_c \approx 1.82$, 
where $T_c = 2/ \ln (1+\sqrt{2}) = 2.269...$ is the
exact critical temperature for the square-lattice Ising model.\cite{ONSA44}  
Having achieved equilibrium with negative magnetization at zero field,
we then subject the system to a constant, uniform, positive magnetic
field, along with a transient heat pulse.  
To simulate the heat pulse, we use a temperature profile given by 
a time-dependent, Gaussian solution of a one-dimensional diffusion equation. 
The profile is centered on the 
mid-line of the Ising lattice, $\bar{x}=63.5$, and each spin in the $x$th column
of the lattice has the temperature
\begin{equation}
T(x,t) = T_0 + 0.3T_c\frac{t_0}{t+t_0}\exp\left( -\frac{(x - 
\bar{x})^2}{4k(t+t_0)}\right), \;\; t \ge 0 \;.
\label{eq:temp}
\end{equation}
Here, $0.3T_c$ is the maximum 
of the temperature pulse, which is attained at $t=0$. Therefore, the peak 
temperature is $T_0 + 0.3T_c = 1.1T_c$.  
The parameter $k$ is the thermal diffusivity, 
which is also set to unity for convenience.  The time $t_0 = \sigma^2/2k$ 
is related to the duration of 
the heat-input process, such that $\sigma$ is the standard deviation that 
governs the width of the temperature profile at $t=0$.\cite{THOM09B} 
Here we use $\sigma=6$ for all simulations. 
(Equation~\ref{eq:temp} most likely underestimates the speed of 
decay of the temperature pulse as it ignores heat conduction into the 
substrate.)
Figure~\ref{fig:pulse} displays the temperature 
of each column  at eight times between $t = 1$ and 500~MCSS.
By first promoting the center-most lattice sites to temperatures above $T_c$ 
before relaxing them back to $T_0$ according to Eq.~(\ref{eq:temp}), we expect 
to initiate a magnetization-switching event that originates along
the center line of the lattice
and propagates outward.  After the completion of this switching process, 
almost all spins will be oriented up, $s_i = +1$.  We define the 
switching time $t_{\rm s}$ as the time until the system first
reaches a magnetization per spin, 
\begin{equation}
m = \frac{1}{L^{2}} \sum_is_i \;,
\label{eq:magn}
\end{equation}
of zero or greater.

\section{RESULTS}
\label{sec-res}

\begin{figure}[ht]
\centering
\begin{tabular}{cc}
\vspace{-40pt}
\epsfig{file=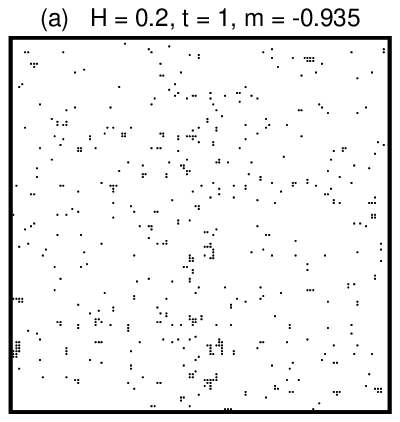,width=0.47\linewidth,clip=} &
\epsfig{file=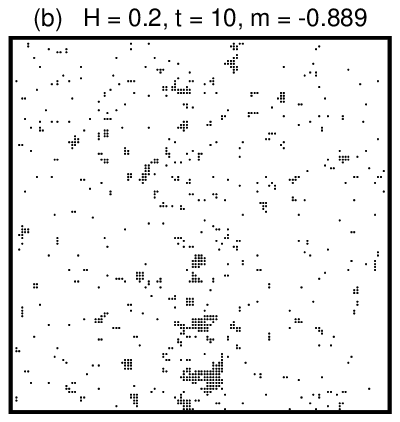,width=0.47\linewidth,clip=}\\ \\ 
\vspace{-40pt}
\epsfig{file=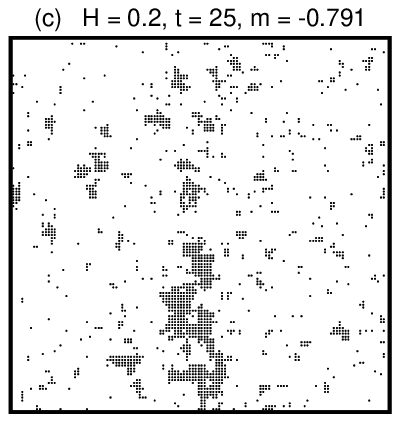,width=0.47\linewidth,clip=} &
\epsfig{file=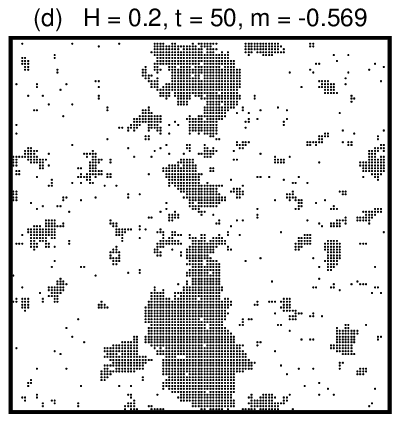,width=0.47\linewidth,clip=}\\ \\
\vspace{-30pt} 
\epsfig{file=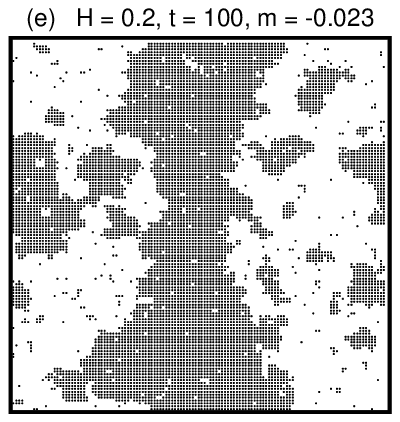,width=0.47\linewidth,clip=} &
\epsfig{file=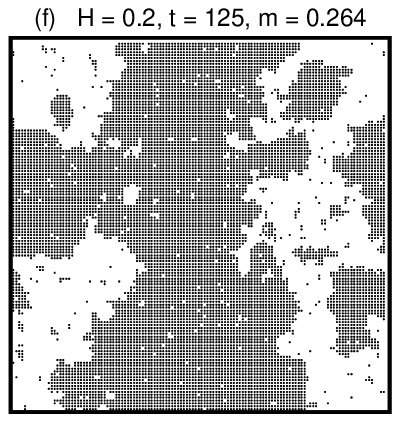,width=0.47\linewidth,clip=}\\ \\ 
\end{tabular} 
\caption{Parts (a)-(f) are snapshots of the $128 \times 128$ Ising system 
at $t = 1$, 10, 25, 50, 
100, and 125~MCSS under influence of the time-dependent temperature profile, 
Eq.~(\ref{eq:temp}), and a constant, uniform applied field of $H = 0.2$. 
Growing clusters of the switched phase are first seen to nucleate near the
center line, where the temperature is highest. However, active nucleation is 
also seen elsewhere in the system.}
\label{fig:snap}
\end{figure}

We first performed a preliminary study 
to confirm that magnetization switching can be induced by 
the temperature profile, given the parameters used in Eq.~(\ref{eq:temp}).  
For this purpose, we inspected snapshots of 
the system during a single run at $H = 0.2$.  In Fig.~\ref{fig:snap} 
we display the configuration 
of the system at six times between $t = 1$ and 125~MCSS
during this run.  As expected, the switching begins near the center line 
of the system, where the temperature 
is above critical, and propagates outward. We note a strong similarity of 
the simulated magnetization configurations to experimental images of ultrathin, 
strongly anisotropic films undergoing magnetization reversal, such as 
Figs.\ 3, 4, and 8 of Ref.~\onlinecite{KIRI97}
and Fig.\ 2 of Ref.~\onlinecite{ROBB08}. This observation further confirms the 
ability of our simplified model to elucidate generic dynamical features 
of real ultrathin films. 

Having confirmed a switching event
at $H=0.2$, statistics were accumulated for 200 simulations at $H=0.2$ and 
also at fifteen weaker fields down to $H=0.06$, 
as detailed in Table~\ref{table}.
For each field, 100 simulations were performed at a constant, uniform 
temperature 
of $T_0 = 0.8T_c$, and 100 were performed using the time-dependent temperature 
profile given by Eq.~(\ref{eq:temp}).  
For each run, the average magnetization for each column at each time step was
recorded along with the switching time, $t_{\rm s}$.

\begin{figure}[ht]
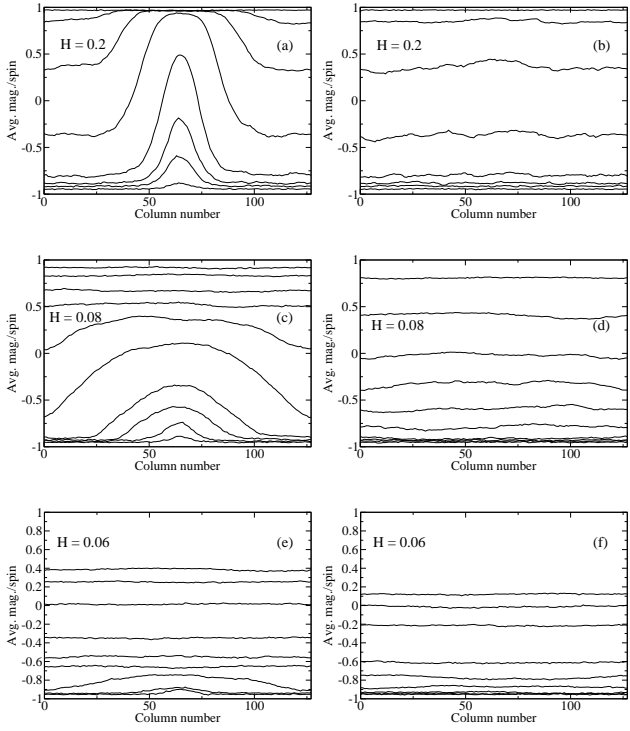

\centering
\begin{tabular}{cc}
\epsfig{file=TempProfile0_2.eps,width=0.47 \linewidth,clip=} &
\epsfig{file=ConstantTemp0_2.eps,width=0.47 \linewidth,clip=} \\ \\ 
\epsfig{file=TempProfile0_08.eps,width=0.47 \linewidth,clip=} &
\epsfig{file=ConstantTemp0_08.eps,width=0.47 \linewidth,clip=} \\ \\ 
\epsfig{file=TempProfile0_06.eps,width=0.47 \linewidth,clip=} &
\epsfig{file=ConstantTemp0_06.eps,width=0.47 \linewidth,clip=} \\ \\ 
\end{tabular}
\caption{The magnetization per spin vs.\ the column number in the lattice, each 
part [(a)--(f)] averaged over 100 independent runs. 
The plots on the left [(a), (c), and (e)] result from the 100 
runs with the relaxing temperature profile, and the ones on the right 
[(b), (d), and (f)] from the 100 runs at a constant, 
uniform temperature, $T_0=0.8T_c$.  
The plots at $H =  0.2$ [(a) and (b)] show the average magnetization per spin 
at $t = 1$, 10, 25, 50, 100, 150, 200, and 300~MCSS from bottom to top.
The plots at $H =  0.08$ [(c) and (d)] show the average magnetization per spin
at $t = 1$, 75, 400, 600, 1000, 1300, 1600, 2000, 3000, and 5500~MCSS 
from bottom to top.  The 
plots at $H = 0.06$  [(e) and (f)] show the average magnetization per spin at 
$t = 1$, 500, 1500, 2500, 4000, 7500, 14000, 20000, and 25000~MCSS. }
\label{fig:mx}
\end{figure}

To investigate the effect that the relaxing 
temperature profile has on each column 
of the Ising lattice,
we plotted the average magnetization per spin against the column number.  
In Fig.~\ref{fig:mx} 
we show this average magnetization for $H = $ 0.2, 0.08, and 0.06. 
The plots on the left [Fig.~\ref{fig:mx}(a), (c), and (e)] 
result from the 100 runs with the relaxing 
temperature profile, and the ones on the 
right [Fig.~\ref{fig:mx}(b), (d), and (f)] from the 100 runs at 
the constant, uniform temperature of $T_0$.  The
plots at $H =  0.2$ [(a) and (b)] show the average magnetization per spin at 
eight different times between $t = 1$ and 300~MCSS.
The plots at $H =  0.08$ [(c) and (d)] show the average magnetization per spin 
at ten different times between $t = 1$ and 5500~MCSS. 
Finally, the plots at $H =  0.06$ [(e) and (f)] show the average magnetization
per spin at nine different times between $t = 1$ and 25000~MCSS. (For a full 
listing of the times, see the figure caption.)
 
\begin{figure}[ht]
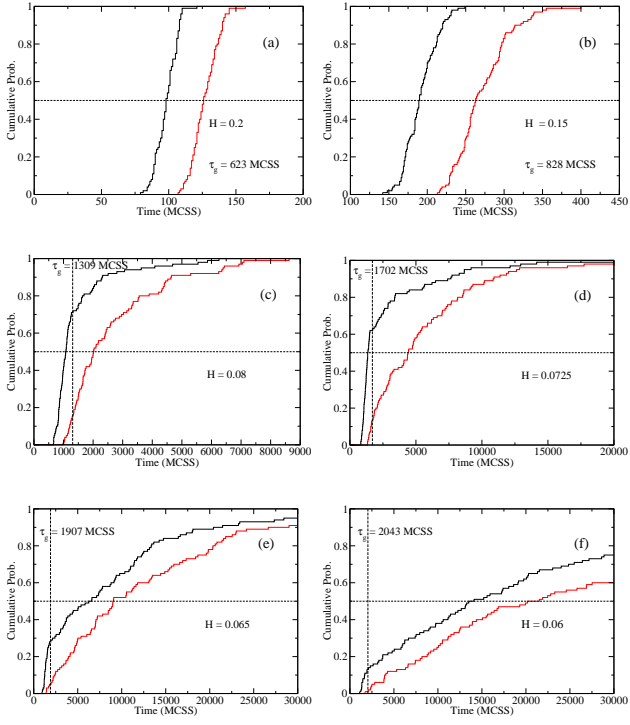

\centering
\begin{tabular}{cc}
\epsfig{file=Fig4a_XMG0_2.eps,width=0.47\linewidth,clip=} &
\epsfig{file=Fig4b_XMG0_15.eps,width=0.47\linewidth,clip=} \\ \\
\epsfig{file=Fig4c_XMG0_08.eps,width=0.47\linewidth,clip=} &
\epsfig{file=Fig4d_XMG0_0725.eps,width=0.47\linewidth,clip=} \\ \\
\epsfig{file=Fig4e_XMG0_065.eps,width=0.47\linewidth,clip=} &
\epsfig{file=Fig4f_XMG0_06.eps,width=0.47\linewidth,clip=} \\ \\ 
\end{tabular}
\caption{Parts (a) -- (f) are cumulative probability distributions for 
the switching times with fields $H = 0.2$, 0.15, 0.08,
0.0725, 0.065, and 0.06, respectively.  The black ``stairs'' correspond to the 
100 simulations with the relaxing
temperature profile (switching times, $t_{\rm s}$).  
The gray (red online) stairs correspond 
to the 100 simulations at uniform temperature (switching times, $t_{\rm c}$). 
The vertical lines in parts (c) -- (f) mark the single-droplet growth time 
$\tau_{\rm g}$.
Note that the time scale increases by more than a factor 100 from (a) to (f).
See discussion in the text.}
\label{fig:cum}
\end{figure}

Again comparing the results with a 
relaxing temperature profile to those realized at 
constant, uniform temperature, in Fig.~\ref{fig:cum}
we show cumulative probability distributions for the switching times for fields 
$H = 0.2$, 0.15, 0.08, 0.0725, 0.065, and 0.06.  The black ``stairs'' are the 
cumulative distributions for the 
switching times in the 100 runs with the relaxing
temperature profile (hereafter referred to as $t_{\rm s}$).  The gray (red 
online) stairs are the cumulative distributions
for the switching times in the 100 runs at constant, uniform temperature 
(hereafter referred to as $t_{\rm c}$).  

Table 1 lists the median switching times for both the 100 runs with the 
relaxing temperature profile ($t_{\rm s}$) and the 100 runs
at constant, uniform temperature $T_0$ ($t_{\rm c}$) for each value of $H$. 
Also listed are the estimated errors $\Delta t_{\rm s}$ and  
$\Delta t_{\rm c}$. 
The last two columns give the ratio $t_{\rm s} / t_{\rm c}$ 
and the associated error $\Delta ( {t_{\rm s}}/{t_{\rm c}} )$.
The error $\Delta t_{\rm s}$ is defined as 
$(t_{{\rm s}2} - t_{{\rm s}1})/2$, 
where $t_{{\rm s}2}$ is the switching time with 
a cumulative probability of $0.55$ and $t_{{\rm s}1}$ is the switching time 
with a cumulative probability of $0.45$, and $\Delta t_{\rm c}$ is defined
analogously.  The error in the ratio $ ( {t_{\rm s}}/{t_{\rm c}} )$ is 
calculated in the standard way as
\begin{equation}
\Delta \left( \frac{t_s}{t_c} \right) = \sqrt{\left( \frac{\Delta t_s}{t_c} 
\right)^2 + \left( \frac{t_s}{t_c^2}\Delta t_c \right)^2} \;.
\label{eq:err}
\end{equation}  
The median switching time has the advantage over the mean 
that it can be estimated even when only half of the 100 simulations 
switch within the maximum number of time steps. This significantly reduces the 
computational requirements, especially for weak fields. 

\begin{figure}[ht]
\centering
\epsfig{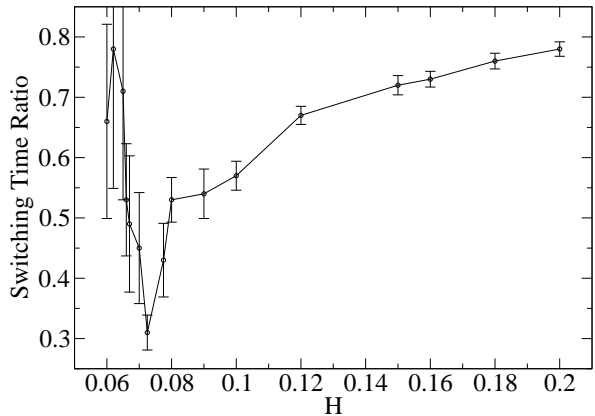}
\caption{The switching-time 
ratio $t_{\rm s}/t_{\rm c}$, shown vs.\ $H$. The minimum value of this 
ratio signifies the maximum benefit from applying the 
relaxing temperature profile of the HAMR method. } 
\label{fig:ratio}
\end{figure}

The ratio $ ( {t_{\rm s}}/{t_{\rm c}} )$ is plotted vs.\ $H$ 
in Fig.~\ref{fig:ratio}. The minimum value of this ratio 
signifies the maximum benefit from using the relaxing 
temperature profile of the HAMR method. The corresponding field  
value, $H=0.0725$, is the optimal field for this simulation. 

\begin{figure}[ht]
\centering
\begin{tabular}{cc}
\vspace{-40pt}
\epsfig{file=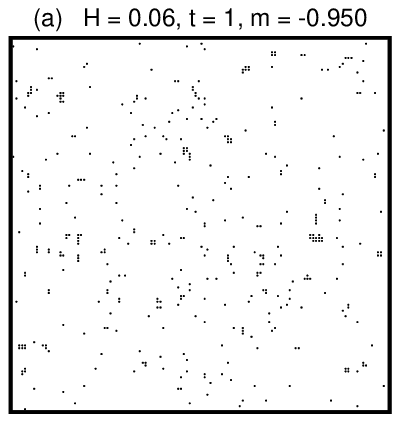,width=0.47\linewidth,clip=} &
\epsfig{file=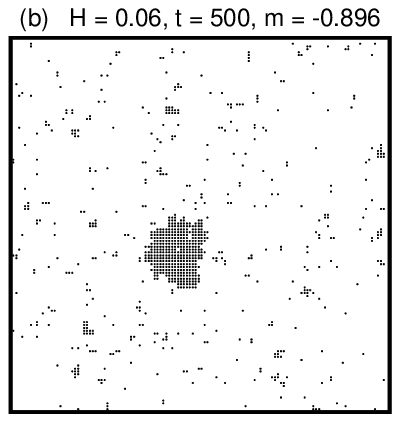,width=0.47\linewidth,clip=}\\ \\ 
\vspace{-40pt}
\epsfig{file=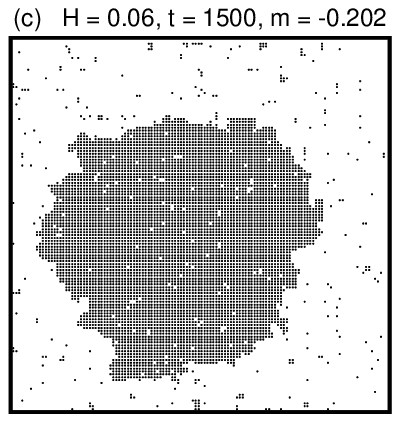,width=0.47\linewidth,clip=} &
\epsfig{file=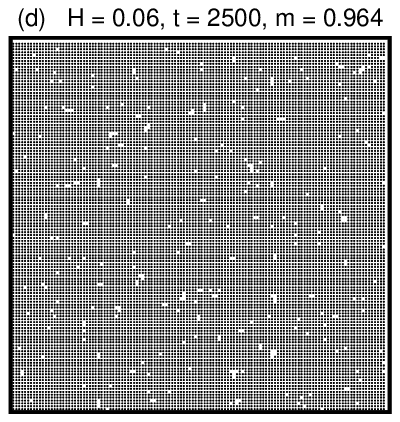,width=0.47\linewidth,clip=}\\ \\
\end{tabular} 
\vspace{5pt}
\caption{Parts (a)-(d) are snapshots of the $128 \times 128$ Ising system 
at $t = 1$, 500, 1500, 
and 2500~MCSS under influence of the time-dependent temperature profile, 
Eq.~(\ref{eq:temp}), and a constant, uniform applied field of $H = 0.06$. 
In this weak field the switching follows the SD mechanism, even in the heated region.
}
\label{fig:snap06A}
\end{figure}

To explain the nonmonotonic shape of the curve representing 
$ ( {t_{\rm s}}/{t_{\rm c}} )$ in Fig.~\ref{fig:ratio}, it is necessary to 
understand the two most important modes of nucleation-initiated 
magnetization switching in finite-sized systems: multidroplet (MD) and 
single-droplet (SD). (For more detailed discussions, 
see Refs.\ \onlinecite{RIKV94A,RIKV94}.) 
The average time between random nucleation events of 
a growing droplet of the equilibrium phase in a $d$-dimensional system of 
linear size $L$ has the strongly field-dependent form, 
$\tau_{\rm n} \propto L^d \exp[\Xi(T) / (T |H^{d-1}|)]$, where $\Xi(T)$ is a 
measure of the free energy associated with the droplet surface.\cite{RIKV94A} 
Once a droplet has nucleated, 
for the weak fields and relatively high temperatures studied in this work
it grows with a near-constant and isotropic radial velocity 
$v_{\rm g} \propto |H|/T$.\cite{RIKV00B} 
As a consequence, the time it would take a newly 
nucleated droplet to grow to fill half of a system of volume of $L^d$ is 
therefore $\tau_{\rm g} \propto L/v_{\rm g}$. 
If $\tau_{\rm g} \gg \tau_{\rm n}$, many droplets will nucleate before the 
first one grows to a size comparable to the system, and many droplets will 
contribute to the switching process. This is the MD regime, which corresponds 
to moderately strong fields and/or large systems.\cite{RIKV94A} 
It is the switching mode shown in Fig.~\ref{fig:snap} for $H=0.2$. 
In the limit of infinitely large systems it 
is identical to the well-known Kolmogorov-Johnson-Mehl-Avrami (KJMA) 
theory of phase transformations.\cite{KOLM37,JOHN39,AVRAMI,RAMO99}  
If $\tau_{\rm g} \ll \tau_{\rm n}$, the first droplet to nucleate will 
switch the system magnetization on its own. This is the SD regime, which 
corresponds to weak fields and/or small systems.\cite{RIKV94A} 
It is the switching mode shown in Fig.~\ref{fig:snap06A} for $H=0.06$.
The crossover region between the SD and MD regimes is known as the Dynamic 
Spinodal (DSP).\cite{RIKV94A} 

\begin{figure}[ht]
\centering
\begin{tabular}{cc}
\vspace{-40pt}
\epsfig{file=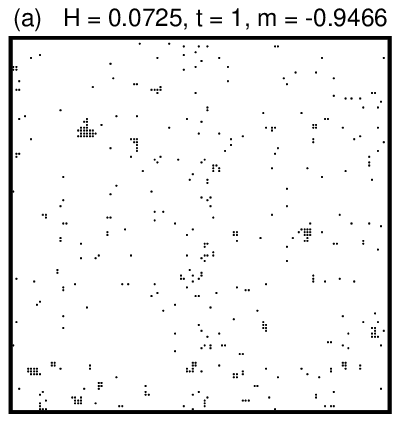,width=0.47\linewidth,clip=} &
\epsfig{file=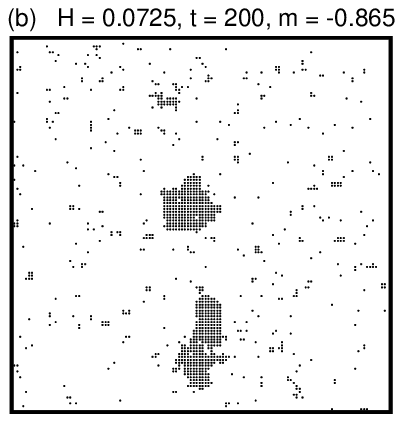,width=0.47\linewidth,clip=}\\ \\ 
\vspace{-40pt}
\epsfig{file=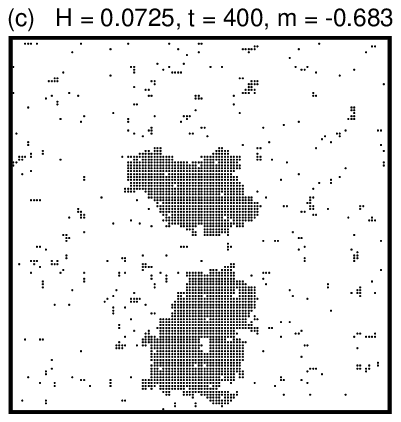,width=0.47\linewidth,clip=} &
\epsfig{file=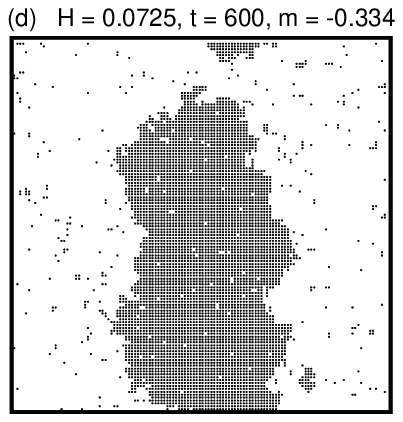,width=0.47\linewidth,clip=}\\ \\
\vspace{-30pt} 
\epsfig{file=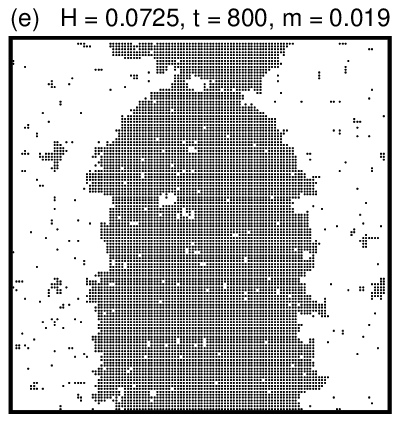,width=0.47\linewidth,clip=} &
\epsfig{file=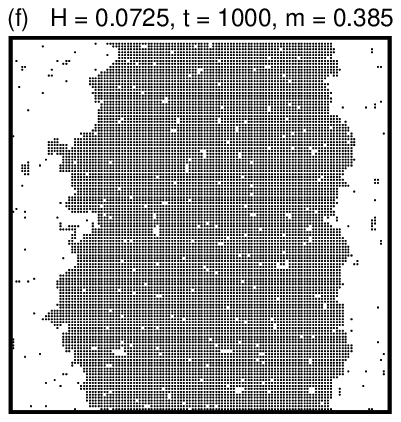,width=0.47\linewidth,clip=}\\ \\ 
\end{tabular} 
\caption{Parts (a)-(f) are snapshots of the $128 \times 128$ Ising system 
at $t = 1$, 200, 400, 600, 
800, and 1000~MCSS under influence of the time-dependent temperature profile, 
Eq.~(\ref{eq:temp}), and a constant, uniform applied field of $H = 0.075$. 
At this intermediate field, multiple 
growing clusters of the switched phase are first seen to nucleate near the
center line, where the temperature is highest. Without the heat pulse, the 
switching would proceed via the SD mechanism.}
\label{fig:snap075}
\end{figure}

One aspect of the MD/SD picture that is particularly relevant to the current 
problem, is the fact that any switching event that takes place at a time 
$t < \tau_{\rm g}$ cannot be accomplished by a single droplet, 
and thus it must be due to the MD mechanism.\cite{BROW01} For a circular 
droplet in a square $L \times L$ system, 
$\tau_{\rm g} \approx L/(\sqrt{2 \pi} v_{\rm g})$. Using results from 
Ref.~\onlinecite{RIKV00B} (which, like the present model, neglects 
pinning effects\cite{KIRI97,ROBB08}), we find that in the range of 
moderately weak fields studied here, at $T = 0.8 T_c$ 
$v_{\rm g}$ can be well approximated as 
$v_{\rm g} \approx 0.75 \tanh{(H/1.82)}$. 
The resulting estimates for $\tau_{\rm g}$ 
in the simulations (which contain {\em no\/} adjustable parameters)  
are shown as vertical lines in Fig.~\ref{fig:cum}(c-f). 
A kink in the cumulative probability distribution for the 
heat-assisted runs is observed at $\tau_{\rm g}$, 
with significantly higher slopes in the MD regime on the short-time 
side of $\tau_{\rm g}$, than in the SD regime on the long-time side.  
From these figures 
we see that the optimal field value for $L=128$, $H = 0.0725$, 
corresponds to the situation where just above 50\% of the heat-assisted switching 
events are caused by the MD mechanism, while essentially all the constant-temperature 
switching events are SD. This situation is illustrated by the series of 
snapshots in Fig.~\ref{fig:snap075}.
For significantly larger fields, both protocols lead to 
all MD switching events [Fig.~\ref{fig:cum}(a,b)], 
while for weaker fields, the great majority of the 
switching events are SD for both protocols
[Fig.~\ref{fig:cum}(e,f)].  In both cases, the ratio 
$t_{\rm s} / t_{\rm c}$ is larger than it is for fields near the 
optimal value [Fig.~\ref{fig:cum}(c,d)]. 
We have  confirmed these conclusions by additional simulations for $L = 64$ and 96 
(not shown).

\section{Conclusions}
\label{sec-conc}

In this paper we have studied a kinetic Ising model of magnetization 
reversal under the influence of a momentary, spatially 
localized input of energy in the form 
of heat (heat-assisted magnetization reversal, or HAMR). Our numerical 
results indicate that the HAMR technique can significantly speed up the 
magnetization reversal in a uniform, applied magnetic field, and we  
find that this speed-up has its optimal 
value at intermediate values of the field. 
This effect is explained in terms of the MD and SD mechanisms of 
nucleation-initiated magnetization switching in finite systems.\cite{RIKV94A} 
The two-dimensional geometry chosen for this study is particularly appropriate 
for thin films. We therefore expect that our predictions 
should be experimentally observable for 
ultrathin ferromagnetic films with strong perpendicular anisotropy, 
such as Co/Pd\cite{CARC85,PURN09} or Co/Pt\cite{ROBB08,FERR02} multilayers. 

\section*{Acknowledgments}

The authors acknowledge useful conversations with M.~A.\ Novotny and 
comments on the manuscript by S.\ von~Moln{\'a}r.  
This work was supported in part by U.S.\ NSF Grants No.\ DMR-0802288 
and DMR-1104829, and 
by the Florida State University Center for Materials Research and Technology 
(MARTECH). 
Computer resources were provided by the Florida State University 
High-performance Computing Center.  




\begin{table}[ht]
\caption{Median 
switching times $t_{\rm s}$ (relaxing temperature profile) 
and $t_{\rm c}$ (constant, uniform temperature) and their ratio
for the sixteen field values included in Fig.~\protect\ref{fig:ratio}.  Also 
given are the estimated errors, $\Delta t_{\rm s}$, $\Delta t_{\rm c}$, 
and $\Delta ( t_{\rm s} / t_{\rm c} )$. 
}
\begin{center}
\begin{tabular}{ccccccc}
\hline  
$H$ & Median $t_{\rm s}$ & $\Delta t_{\rm s}$ 
  & Median $t_{\rm c}$ & $\Delta t_{\rm c}$ & $( t_{\rm s}/t_{\rm c} )$ 
&  $\Delta ( t_{\rm s}/t_{\rm c} )$   \\
\hline
0.2000 & 98.0 & 1.0  & 126.0 & 1.5 & 0.778 &  0.012\\
0.1800 & 126.0 & 1.5  & 166.5 & 2.0 & 0.757 & 0.013 \\
0.1600 & 165.0 & 0.5 & 225.5 & 4.0 & 0.732 & 0.013 \\
0.1500 & 189.5 & 2.5 & 263.0 & 4.5  & 0.721 & 0.016  \\
0.1200 & 323.0 &6.0 & 482.5 & 6.5 & 0.669&  0.015 \\
0.1000 & 504.0 & 16.0 & 881.5 & 25.0 & 0.572 & 0.024 \\
0.0900 & 670.5 & 14.5 & 1253.0 & 92.0 & 0.535 & 0.041  \\
0.0800 & 1077.0 & 43.0 & 2015.0 & 113.0  & 0.534 &  0.037 \\
0.0775& 1148.0 & 72.0 & 2676.0 & 344.5 & 0.429 & 0.061 \\
0.0725& 1374.0 & 63.5 & 4413.0 & 354.5 & 0.311 & 0.029\\
0.0700 & 2443.0 & 274.5 & 5470.0 & 938.0 & 0.447& 0.092  \\
0.0670 & 3232.0 & 629.5 & 6621.5 & 838.5 & 0.488& 0.113\\
0.0660 & 4035.5 & 580.5 & 7562.5 & 733.0 & 0.534& 0.093 \\
0.0650 & 6426.5 & 1453.0 & 9030.0 & 1035.5 & 0.712 & 0.180\\
0.0620 & 11569.5 & 1927.0 & 14788.0 & 3607.0 & 0.782& 0.231 \\
0.0600 & 13808.0 & 2479.0 & 20851.0 & 3435.5 & 0.662 & 0.161\\
\hline

\end{tabular}
\end{center}
\label{table}
\end{table}


\newpage

\vspace{40pt}

\end{document}